\documentclass[aps,preprint]{revtex4}%
\usepackage{amsfonts}
\usepackage{amsmath}
\usepackage{amssymb}
\usepackage{graphicx}
\usepackage{pdfpages}
\usepackage{xcolor}
\usepackage{caption}
\usepackage{lineno,hyperref}
\usepackage{changes}%
\usepackage{subfig}
\providecommand{\U}[1]{\protect\rule{.1in}{.1in}}

\begin{document}
\preprint{HEP/123-qed}
\begin{center}
{\Large The Dirac and Klein-Gordon equations and Greybody Radiation for the Regular Hayward Black Hole}

Ahmad Al-Badawi

Department of Physics, Al-Hussein Bin Talal University, P. O. Box: 20,
71111, Ma'an, Jordan

\bigskip E-mail: ahmadbadawi@ahu.edu.jo

{\Large Abstract}
\end{center}

We investigate the Dirac and Klein-Gordon equations, as well as greybody radiation, for the Hayward black hole (BH) spacetime. We first consider the Dirac equation using a null tetrad in the Newman- Penrose (NP) formalism. The  equations are then separated into angular and radial parts. A pair of one-dimensional Schrödinger like wave equations with effective potentials is obtained from the radial part. In order to examine the behavior of potentials, they are plotted with respect to radial distances. Additionally, the Klein-Gordon equation is considered in the  Hayward BH spacetime. At the end, we compute greybody factors for bosons and fermions and our results are shown graphically and discussed.
\volumeyear{ }
\eid{ }
\date{\today}
\received{}

\maketitle
\tableofcontents

\section{Introduction}

Regular black holes (BHs) were discovered first by Bardeen and came to be known as Bardeen BH [1-3]. These types of BH do have an event horizon but no spacetime singularities. The singularity theorem is bypassed in such a way that these BH solutions only satisfy the weak energy condition
and not the strong one. Although a regular BH is not the solution to the Einstein vacuum field equations, it can be obtained either by using nonlinear
electrodynamics; or by reinventing gravity. The charged version of Bardeen BH was derived by the coupling of Einstein and Maxwell field equations by Ayon-Beato and Garcia [4].  They introduced a nonlinear electric field as a source of charge for the solution of Maxwell field equations. Bronnikov [5-7] later demonstrated analytically that charged regular BHs are not correct solutions of the field equations because the electromagnetic Lagrangian in these solutions is invariably different in various portions of space.

Hayward [8], on the other hand, proposed a regular (non singular) BH solution that is comparable to Bardeen BH and has no charge term. This BH solution consists of a compact spacetime region of trapped surfaces with circularly joined inner and outer bounds as a single smooth trapping horizon [8]. The Hayward BH has well-defined asymptotic limits, such as in the limit $r\rightarrow \infty $, it is Schwarzschild BH and \ de-Sitter BH $%
r\rightarrow 0$. The regular Hayward BH has been the subject of a lot of research [9-14]. The rotating and modified BH of Hayward is discussed in [15-16], along with its implications. Recently, the quasinormal frequencies for a class of regular black hole solutions generalizing Bardeen and Hayward spacetimes were studied by considering the sixth order WKB approximation [17]. In addition, Hayward spacetime has also been studied by solving Maxwell's equations for Hayward geometry and considering a Hayward spacetime immersed in a magnetic field [18].

In BH physics, regular BHs are especially interesting since they lack physical singularities. This is the fundamental impetus for our current research. Investigating scalar/spinor fields, as well as its characteristic bosonic and fermionic quantum radiation, is a secondary incentive for better understanding the Hayward BH. A seminal investigation of perturbations and greybody radiations in various gravitational spacetimes has recently been studied [19-31].

The organization of the paper is as follows. Sect. 2 discusses the Hayward BH, writes the Dirac equation in Hayward spacetime, separates the equation into angular and radial parts, and examines the radial equations to end up with a pair of wave equations with effective potentials.  In sect. 3, we study the Klein-Gordon equation in Hayward BH spacetime, and we calculate the greybody factors of Hayward BH spacetime for fermions as well as bosons. The paper ends with the conclusion in Sect. 4.

\section{Dirac Equation in Hayward BH Spacetime}

The spherically symmetric static Hayward non singular BH is given by the
following line element [8]:

\begin{equation}
ds^{2}=f\left( r\right) dt^{2}-f^{-1}(r)dr^{2}-r^{2}\left( d\theta ^{2}+\sin
^{2}\theta d\phi ^{2}\right)  \label{3}
\end{equation}%
where the lapse function $f\left( r\right) $ has the following form
\begin{equation}
f\left( r\right) =1-\frac{2Mr^{2}}{r^{3}+2Ml^{2}}.  \label{4}
\end{equation}%
in which $M$ is the mass of the BH and $l$\ is the Hayward parameter. The spacetime metric  (\ref{3}) admits three kinds of different spacetime: no horizon $\left( 
\frac{l^{2}}{M^{2}}>\frac{16}{27}\right) ,$ one horizon $\left( \frac{l^{2}}{%
M^{2}}=\frac{16}{27}\right) $ and two horizons $\left( \frac{l^{2}}{M^{2}}<%
\frac{16}{27}\right)$ as shown in Fig. (1). The lapse function behaves as%
\begin{eqnarray}
\lim_{r\rightarrow \infty }f\left( r\right) &=&1-\frac{2M}{r}+O\left( \frac{1%
}{r^{4}}\right) \rightarrow \text{Schwarzschild BH} \\
\lim_{r\rightarrow 0}f\left( r\right) &=&1-\frac{r^{2}}{l^{2}}+O\left(
r^{5}\right) \rightarrow \text{de-Sitter BH.}  \notag
\end{eqnarray}%
As a result, at large values of $r$, the Hayward non singular BH becomes a Schwarzschild BH, while at small values of $r$, it becomes a de-Sitter BH. 
\begin{figure}[h]
\centering
\includegraphics[scale=1]{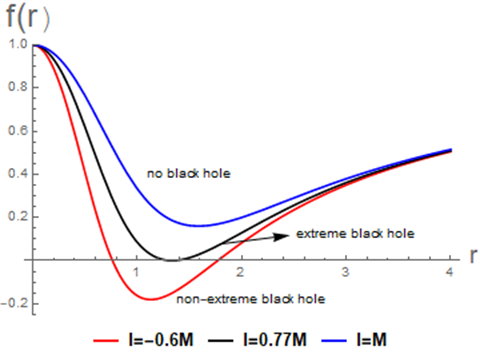} \caption{ Horizons of the Hayward black hole spacetime with different values of $l$}
\label{Figure1}%
\end{figure}
We use the Newman- Penrose (NP) formalism [32] to express and solve the Dirac equation in spacetime metric (\ref{3}). Therefore, let us define the
complex null tetrad vectors $\left( l,n,m,\overline{m}\right) $ for the
metric (\ref{3}) where they satisfy the orthogonality conditions, $\left(
l.n=-m\text{.}\overline{m}=1\right) $ as%
\begin{equation*}
l_{\mu }=\left( 1,-\frac{1}{f\left( r\right) },0,0\right) ,
\end{equation*}%
\begin{equation*}
n_{\mu }=\frac{1}{2}\left( f\left( r\right) ,1,0,0\right) ,
\end{equation*}%
\begin{equation*}
m_{\mu }=\frac{-r}{\sqrt{2}}(0,0,1,i\sin \theta ),
\end{equation*}%
\begin{equation}
\overline{m}_{\mu }=\frac{-r}{\sqrt{2}}(0,0,1,-i\sin \theta ),  \label{6}
\end{equation}%
where a bar over a quantity denotes complex conjugation and the dual
co-tetrad are given by 
\begin{equation*}
l^{\mu }=\left( \frac{1}{f\left( r\right) },1,0,0\right) ,
\end{equation*}%
\begin{equation*}
n^{\mu }=\frac{1}{2}\left( 1,-f\left( r\right) ,0,0\right) ,
\end{equation*}%
\begin{equation*}
m^{\mu }=\frac{1}{\sqrt{2}r}(0,0,1,\frac{i}{\sin \theta }),
\end{equation*}%
\begin{equation}
\overline{m}^{\mu }=\frac{1}{\sqrt{2}r}(0,0,1,\frac{-i}{\sin \theta }).
\label{7}
\end{equation}%
The non-zero spin coefficients can then be computed as 
\begin{align}
\rho & =-\frac{1}{r},\mu =-\frac{f}{2r},\qquad  \notag \\
\gamma & =\frac{f'}{4} ,\beta =-\alpha =\frac{\cot \theta }{2%
\sqrt{2}r}.  \label{8}
\end{align}

Now, we shall employ the Chandrasekar-Dirac equations (CDEs) [33]
\begin{equation*}
\left( D+\epsilon -\rho \right) F_{1}+\left( \overline{\delta }+\pi -\alpha
\right) F_{2}=i\mu _{0}G_{1},
\end{equation*}%
\begin{equation*}
\left( \Delta +\mu -\gamma \right) F_{2}+\left( \delta +\beta -\tau \right)
F_{1}=i\mu _{0}G_{2},
\end{equation*}%
\begin{equation*}
\left( D+\overline{\epsilon }-\overline{\rho }\right) G_{2}-\left( \delta +%
\overline{\pi }-\overline{\alpha }\right) G_{1}=i\mu _{0}F_{2},
\end{equation*}%
\begin{equation}
\left( \Delta +\overline{\mu }-\overline{\gamma }\right) G_{1}-\left( 
\overline{\delta }+\overline{\beta }-\overline{\tau }\right) G_{2}=i\mu
_{0}F_{1},  \label{12n}
\end{equation}%
where $F_{1},F_{2},G_{1}$and $G_{2}$ represent the components of the wave
function \textquotedblright Dirac spinors\textquotedblright\ and $\mu _{0}=%
\sqrt{2}\mu _{p}$ is the mass of the particle. The directional derivatives
in CDEs are defined by $D=l^{\mu }\partial _{\mu },\Delta =n^{\mu }\partial
_{\mu }$ and $\delta =m^{\mu }\partial _{\mu }$.

Substituting the spin coefficients and the directional derivatives into CDEs leads to
\begin{equation*}
\left( \emph{D}-\frac{1}{r}\right) F_{1}+\frac{1}{\sqrt{2}r}\mathit{L}%
F_{2}=i\mu _{0}G_{1},
\end{equation*}%
\begin{equation*}
\frac{-f}{2}\left( \mathit{D}^{\dag }-\frac{f^{\prime }}{2f}+\frac{1}{r}%
\right) F_{2}+\frac{1}{\sqrt{2}r}\mathit{L}^{\dag }F_{1}=i\mu _{0}G_{2},
\end{equation*}%
\begin{equation*}
\left( \emph{D}+\frac{1}{r}\right) G_{2}-\frac{1}{\sqrt{2}r}\mathit{L}^{\dag
}G_{1}=i\mu _{0}F_{2},
\end{equation*}%
\begin{equation}
\frac{f}{2}\left( \mathit{D}^{\dag }-\frac{f^{\prime }}{2f}+\frac{1}{r}%
\right) G_{1}+\frac{1}{\sqrt{2}r}\mathit{L}G_{2}=i\mu _{0}F_{1},  \label{D10}
\end{equation}%
where the operators are 
\begin{equation*}
\mathit{D}^{\dag }=-\frac{2}{f}\Delta,
\end{equation*}
\begin{equation*}
\mathit{L}=\sqrt{2}r\overline{\delta }+\frac{\cot \theta }{2},
\end{equation*}%
\begin{equation}
\mathit{L}^{\dag }=\sqrt{2}r\delta +\frac{\cot \theta }{2}.
\end{equation}

We assume a separable solution in the form of $%
F=F\left( r,\theta \right) \exp \left( ikt\right)
\exp \left( im\phi \right)$, where $k$ is the
frequency of the incoming wave and $m$ is the azimuthal quantum number of the wave: 
\begin{equation}
F_{1}=R_{1}\left( r\right) A_{1}\left( \theta \right) \exp \left( ikt\right)
\exp \left( im\phi \right) ,  \notag
\end{equation}%
\begin{equation}
F_{2}=R_{2}\left( r\right) A_{2}\left( \theta \right) \exp \left( ikt\right)
\exp \left( im\phi \right) ,  \notag
\end{equation}%
\begin{equation}
G_{1}=R_{2}\left( r\right) A_{1}\left( \theta \right) \exp \left( ikt\right)
\exp \left( im\phi \right) ,  \notag
\end{equation}%
\begin{equation}
G_{2}=R_{1}\left( r\right) A_{2}\left( \theta \right) \exp \left( ikt\right)
\exp \left( im\phi \right) .  \label{10}
\end{equation}%
Substituting Eqs. (\ref{10}) into Eq.s (\ref{D10}), the CDEs transform into 
\begin{equation*}
\frac{r}{R_{2}}\left( \emph{D}-\frac{1}{r}\right) R_{1}+\frac{1}{\sqrt{2}%
A_{1}}\mathit{L}A_{2}=ir\mu _{0},
\end{equation*}%
\begin{equation*}
\frac{-rf}{2R_{1}}\left( \mathit{D}^{\dag }-\frac{f^{\prime }}{2f}+\frac{1}{r%
}\right) R_{2}+\frac{1}{\sqrt{2}A_{2}}\mathit{L}^{\dag }A_{1}=ir\mu _{0},
\end{equation*}%
\begin{equation*}
\frac{r}{R_{2}}\left( \emph{D}+\frac{1}{r}\right) R_{1}-\frac{1}{\sqrt{2}%
A_{2}}\mathit{L}^{\dag }A_{1}=ir\mu _{0},
\end{equation*}%
\begin{equation}
\frac{rf}{2R_{1}}\left( \mathit{D}^{\dag }-\frac{f^{\prime }}{2f}+\frac{1}{r}%
\right) R_{2}+\frac{1}{\sqrt{2}A_{1}}\mathit{L}A_{2}=ir\mu _{0}.
\end{equation}%
In order to separate the equations, we define a separation constant. This is
carried out by using the angular equations. Actually, it is already known
from the literature that the separation constant can be expressed in terms
of the spin-weighted spheroidal harmonics. The radial parts of CDEs become%
\begin{equation}
\left( \emph{D}+\frac{1}{r}\right) R_{1}=\frac{1}{r}\left( \lambda +ir\mu
_{0}\right) R_{2},  \label{f11}
\end{equation}
\begin{equation}
\frac{f}{2}\left( \mathit{D}^{\dag }+\frac{f^{\prime }}{2f}+\frac{1}{r}%
\right) R_{2}=\frac{1}{r}\left( \lambda -ir\mu _{0}\right) R_{1}.
\label{f12}
\end{equation}%
where $\lambda$ is the separation constant. We further assume that $R_{1}\left( r\right) =\frac{P_{1}\left( r\right) }{r},R_{2}\left( r\right) =%
\frac{P_{2}\left( r\right) }{r}$
, then Eqs. (\ref{f11}, \ref{f12}) transform into, 
\begin{equation}
\emph{D}P_{1}=\frac{1}{r}\left( \lambda +ir\mu _{0}\right) P_{2},
\label{f13}
\end{equation}
\begin{equation}
\frac{f}{2}\left( \mathit{D}^{\dag }+\frac{f^{\prime }}{2f}\right) P_{2}=%
\frac{1}{r}\left( \lambda -ir\mu _{0}\right) P_{1}.  \label{f14}
\end{equation}%
Using the new functions $P_{1}\left( r\right) =T_{1}\left( r\right) ,P_{2}\left( r\right) =\sqrt{%
\frac{f}{2}}T_{2}\left( r\right)$, 
and defining the tortoise coordinate $r_{\ast }$ as $\frac{d}{dr_{\ast }}=f%
\frac{d}{dr}$, Eqs. (\ref{f13}, \ref{f14}) become 
\begin{equation}
\left( \frac{d}{dr_{\ast }}+ik\right) T_{1}=\sqrt{\frac{f}{r}}\left( \lambda
+ir\mu _{0}\right) T_{2},  \label{f15}
\end{equation}

\begin{equation}
\left( \frac{d}{dr_{\ast }}-ik\right) T_{2}=\sqrt{\frac{f}{r}}\left( \lambda
-ir\mu _{0}\right) T_{1}.  \label{f16}
\end{equation}%
To write Eqs. (\ref{f15}, \ref{f16}) in a more compact form, we combine the
solutions in the following way: $\psi _{+}=T_{1}+T_{2},\psi _{-}=T_{2}-T_{1}$
. Hence, we end up with a pair of one-dimensional Schr\"{o}dinger like
equations with effective potentials,%
\begin{equation}
\frac{d^{2}\psi _{+}}{dr_{\ast }^{2}}+\left( k^{2}-V_{+}\right) \psi _{+}=0,
\end{equation}%
\begin{equation}
\frac{d^{2}\psi _{-}}{dr_{\ast }^{2}}+\left( k^{2}-V_{-}\right) \psi _{-}=0,
\end{equation}%
\begin{equation}
V_{\pm } =\frac{r^{2}B^{3}f}{D^{2}}\pm \frac{rB^{3/2}}{D^{2}}%
\sqrt{f}\left( \left( r-M\right) B+3r^{3}\mu _{0}-\frac{6r^{5}\mu _{0}M}{%
r^{3}+2Ml^{2}}\right)  \label{p1}
\end{equation}
\begin{equation}
\mp \frac{r^{3}B^{5/2}}{D^{3}}f^{3/2}\left( \left( 2rB+2r^{3}\mu
_{0}\right) +\frac{\left( r-M\right) \lambda \mu _{0}}{k}\right)  \notag
\end{equation}

where 
\begin{equation}
B=\left( \lambda ^{2}+\mu _{0}r^{2}\right) ,\qquad D=r^{2}B+\frac{\lambda
\mu _{0}r^{2}}{2k}\left( 1-\frac{2Mr^{2}}{r^{3}+2Ml^{2}}\right) .
\end{equation}%
The massless Dirac particle (fermions) can be obtained by setting $%
\mu
_{0}=0$ in (\ref{p1}) namely

\begin{equation}
V_{\pm }=\frac{\lambda ^{2}}{r^{2}}f\pm \frac{\lambda \left( r-M\right) }{%
r^{3}}\sqrt{f}\mp \frac{2\lambda }{r^{2}}f^{3/2}.  \label{p2}
\end{equation}
Two-dimensional plots are used to observe the Hayward parameter effect and to comprehend the physical behavior of potentials in a physical region. The behavior of the effective potentials (\ref{p1}) for massive particles is shown in figures 2 and 3 for various values of the Hayward parameter. From the behaviour showed in 2 and 3, we can see that there are modest peaks in the physical region for sufficiently large Hayward parameters. In addition, we can note that the Schwarzschild  case $(l=0)$ has the lowest potential. Regardless of the Hayward parameter's value, potentials become bounded and approach a constant value for large $r$ values. We notice that the behavior of the potentials in the massless case $( \mu_{0}=0)$ is similar to that shown in Figures 2 and 3. 
\begin{figure}[h]
\centering
\includegraphics[scale=1]{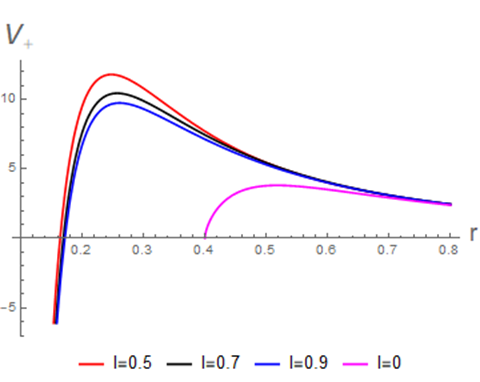} \caption{Effective potential $V_{+}$ (\ref{p1}) with respect to the coordinate r. Here, $M =0.2, k = 0.4, \mu = 0.1,$ and $\lambda =2$.}
\label{Figure1}%
\end{figure}
\begin{figure}[h]
\centering
\includegraphics[scale=1]{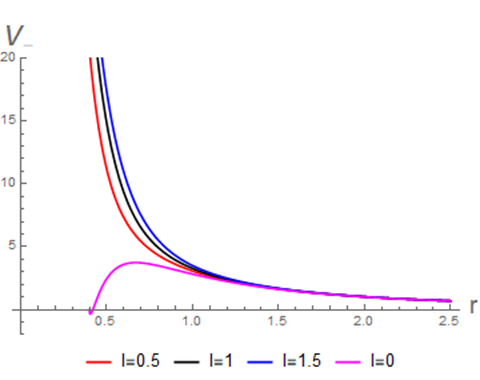} \caption{Effective potential $V_{-}$ (\ref{p1}) with respect to the coordinate r. Here, $ k = 0.2, \mu = 0.1,M=0.4$ and $\lambda =1$.}
\label{Figure1}%
\end{figure}
\section{Greybody Radiation from Hayward BH}

\subsection{Klein-Gordon equation}

The massless Klein-Gordon equation of scalar field in spherical coordinates is written as 
\begin{equation}
\frac{1}{\sqrt{-g}}\partial _{\mu }\sqrt{-g}g^{\mu \nu }\partial _{\nu }\Psi
(t,r,\theta ,\phi )=0,  \label{is1}
\end{equation}

where $g$ is the determinant of the spacetime metric (\ref{3}), so that $%
\sqrt{-g}=r^{2}\sin \theta $. For the metric (\ref{3}) the Klein-Gordon
equation becomes 
\begin{equation}
\frac{1}{f}\frac{\partial ^{2}\Psi }{\partial t^{2}}+\frac{1}{r^{2}}\frac{%
\partial }{\partial r}\left( r^{2}f\frac{\partial }{\partial r}\right) \Psi +%
\frac{1}{r^{2}\sin \theta }\frac{\partial }{\partial \theta }\left( \sin
\theta \frac{\partial }{\partial \theta }\right) \Psi +\frac{1}{r^{2}\sin
^{2}\theta }\frac{\partial ^{2}\Psi }{\partial \phi ^{2}}=0.  \label{is2}
\end{equation}%
To separate the variables in (\ref{is2}), we assume the solutions to the
wave equation in the form 
\begin{equation}
\Psi =R\left( r\right) \Theta \left( \theta \right) \exp \left( -i\omega
t\right) \exp \left( im\phi \right) ,
\end{equation}

where $\omega $ denotes the frequency of the wave and $m$ is the azimuthal
quantum number. Therefore, the radial part of Klein-Gordon equation (\ref%
{is2}) becomes%
\begin{equation}
\frac{d}{dr}\left( r^{2}f\frac{d}{dr}\right) R\left( r\right) +\left( \frac{%
r^{2}}{f}\omega ^{2}-\lambda \right) R\left( r\right) =0,  \label{is5}
\end{equation}%
where $\lambda $ is the eigenvalue coming from the physical solution of the
angular equation of $\Theta \left( \theta \right) $, namely the standard
spherical harmonics [34]. By changing the variable in a new form as $R\left( r\right) =\frac{U\left(
r\right) }{r},$ the radial wave equation (\ref{is5}) recasts into a one
dimensional Schr\"{o}dinger like equation as follows%
\begin{equation}
\frac{d^{2}U}{dr_{\ast }^{2}}+\left( \omega ^{2}-V_{eff}\right) U=0,
\end{equation}%
in which $r_{\ast }$ is the tortoise coordinate: $\frac{dr_{\ast }}{dr}=%
\frac{1}{f},$ and $V_{eff}$ is the effective potential given by
\begin{equation}
V_{eff}=\left(1-\frac{2Mr^{2}}{r^{3}+2Ml^{2}}\right)\left( \frac{\lambda }{r^{2}}+\frac{2Mr^{3}-8M^{2}l^{2}}{({r^{3}+2Ml^{2})^{2}}}\right) .
\label{v1}
\end{equation}

Based on the above equation, the effective potential (\ref{v1}) is dependent on Hayward parameters. Recall that Fig. 1 shows that Hayward spacetimes can have two horizons, which correspond to the BH spacetimes, one horizon, which corresponds to the extremal BH spacetimes, and no horizons, which correspond to no-horizon spacetimes, depending on the mass $M$ and Hayward parameter $l$. The effective potential vanishes at the two horizons, as the lapse function also vanishes there. Moreover, the lapse function in metric (\ref{3})  governing the event horizon location, ($r$ approaches the event horizon as $r_{\ast }$ approaches $-\infty$). Hence, the effective potential will be restricted here only to the regions located outside the event horizon. Figure 5 illustrates how the Hayward parameter affects the effective potential. It indicates that with Hayward parameter, the potentials become more stable than with no Hayward parameter $(l=0)$, implying that their primary function is to formalize potentials and reduce their peaks. As a result, we can deduce that when the potential peaks decrease, the greybody factor will increase.
\begin{figure}[h]
\centering
\includegraphics[scale=1]{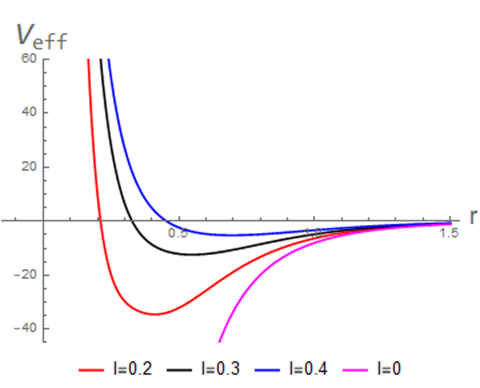} \caption{Dependence of potential (\ref{v1}) on radial coordinate $r$ for several values of Hayward parameter $l$. Here, $\lambda =6 $ and $ M = 1$.}
\label{Figure1}%
\end{figure}

\subsection{Greybody Factor of bosons}

The greybody factor is used to calculate the deviation of thermal spectra from the black body or Hawking radiation [35, 36]. In essence, the greybody factor is a physical quantity related to the quantum nature of a BH. A high greybody factor value indicates a high probability that Hawking radiation can reach spatial infinity [37, 38]. In this section, we assume that Hawking radiation is a massless scalar field that obeys the Klein-Gordon equation for bosons and the Dirac equation for fermions. We will use the general method of semi-analytical bounds for the greybody factors [39,40] to compute the greybody factor of the Hayward BH spacetime as follows, 
\begin{equation}
\sigma _{l}\left( \omega \right) \geq \sec h^{2}\left( \int_{-\infty
}^{+\infty }\wp dr_{\ast }\right) ,  \label{is8}
\end{equation}

in which $r_{\ast }$ is the tortoise coordinate and 
\begin{equation}
\wp =\frac{1}{2h}\sqrt{\left( \frac{dh\left( r_{\ast }\right) }{dr_{\ast }}%
\right) ^{2}+(\omega ^{2}-V_{eff}-h^{2}\left( r_{\ast }\right) )^{2}},
\label{is9}
\end{equation}

where $h(r\ast )$ is a positive function satisfying $h\left( -\infty \right)
=h\left( -\infty \right) =\omega $. For more details, one can see [39, 40]. We
select $h=\omega $. Thus, Eq. \ref{is9} becomes
\begin{equation}
\sigma \left( \omega \right) \geq \sec h^{2}\left( \int_{r_{h}}^{+\infty }%
\frac{V_{eff}}{2\omega }dr_{\ast }\right) .  \label{gb1}
\end{equation}%
The greybody factor for bosons is computed using the potential from (\ref{v1}). However, we use the classical term-by-term integration technique for obtaining asymptotic integral expansions [41, 42], so (\ref{gb1}) becomes
\begin{equation}
\sigma \left( \omega \right) \geq \sec h^{2}\frac{1}{2\omega }\left( \frac{%
\lambda }{r_{h}}+\frac{M}{r_{h}^{2}}-\frac{16M^{2}l^{2}}{5r_{h}^{5}}+\frac{%
M^{2}l^{4}\left( 3+4M\right) }{r_{h}^{8}}\right) .  \label{gb2}
\end{equation}
We plot $\sigma \left( \omega \right) $ with various values of the Hayward parameter in Figure 5 to see how this bosonic greybody factor behaves. It can be seen that the Hayward parameter suppresses the greybody factor.
\begin{figure}[h]
\centering
\includegraphics[scale=1]{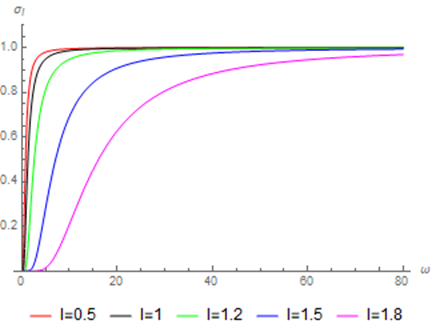} \caption{Plots of $\sigma \left( \omega \right)$ (\ref{gb1}) versus $\omega$ for bosons. The physical parameters are chosen as; $M = r_{h} = \lambda = 1$.}
\label{Figure1}%
\end{figure}

\subsection{Greybody Factors of fermions}

The fermionic greybody factor of the neutrinos emitted from the Hayward BH will be computed here. To this objective, the potentials (\ref{p2}) are considered. Using the approach given above [see Eqs. (\ref{is8})-(\ref%
{gb1})], we get 
\begin{equation}
\sigma \left( \omega \right) \geq \sec h^{2}\frac{\lambda }{2\omega }\left( 
\frac{\lambda +1}{r_{h}}\mp \frac{M}{r_{h}^{2}}\left( 1+\frac{M}{2r_{h}}+%
\frac{M^{2}}{2r_{h}^{2}}+\frac{25M^{3}-48Ml^{2}}{40r_{h}^{3}}\right) \right.
\label{gb3}
\end{equation}%
\begin{equation*}
\left. \pm \frac{7M^{5}+48M^{3}l^{2}}{16r_{h}^{6}}\right).
\end{equation*}
\begin{figure}[h]
\centering
\includegraphics[scale=1]{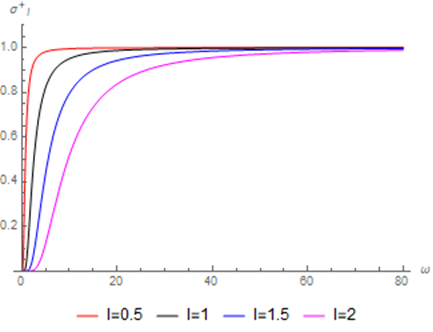}\caption{Plots of $\sigma_{+}\left(\omega \right)$ (\ref{gb3}) versus $\omega$ for fermions with spin-(+1/2). The physical parameters are chosen as; $\lambda $ = $M =r_{h}= 1$.}
\label{FigureK2}%
\end{figure}
\begin{figure}[h]
\centering
\includegraphics[scale=1]{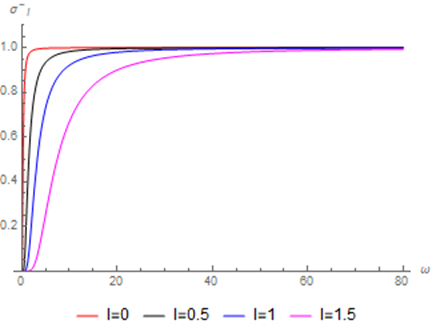} \caption{Plots of $\sigma_{-}\left(\omega \right)$ (\ref{gb3})  versus $\omega$ for fermions with spin-(-1/2). The physical parameters are chosen as; $\lambda $ = $M =r_{h}= 1$.}
\label{FigureK2}%
\end{figure}
The behavior of spin-(+1/2) and spin-(-1/2) under the effect of the Hayward parameter is depicted in Figures 6 and 7. The figures indicate that, like with bosons, when the Hayward parameter $l$ grows, the greybody factor drops.
\section{Conclusion}

In this work, we investigated the Dirac and Klein-Gordon equations, as well as greybody radiation in Hayward spacetime, which describes a non-singular BH. In section II, we used null tetrads in the NP formalism to investigate the Dirac equation in a Hayward spacetime background. The Dirac equations have been separated into radial and angular sets. The radial equations are investigated, and a pair of one-dimensional Schrödinger-like equations with effective potentials are obtained. The effective potentials are affected by both the mass and the Hayward parameter.
In order to understand the physical interpretations, we plot the potentials (\ref{p1}) for various Hayward parameter values. It is shown that, the potential barriers are higher for small values of the Hayward parameter, as illustrated in Figs. 2 and 3. When the Hayward parameter is reduced, the peak of the potential barriers increases noticeably. Also, the potential tends to a constant value for large distances regardless of the Hayward parameter.

In section III, we studied the Klein-Gordon equation in Hayward spacetime and reduced the radial part to one-dimensional Schrödinger equations with effective potentials. Furthermore, we looked into the effect of the Hayward parameter on the effective potentials (Fig. 4). In our study, we found that the potential become more stable when the Hayward parameter was reduced. We also calculated an analytical solution for the greybody factor, which is one of the fundamental pieces of information that can be obtained from BHs for both bosons and fermions. To analyze the effect of the Hayward parameter, the greybody factors in Hayward BH are graphically illustrated. Figures 5, 6, and 7 show that as the Hayward parameter value increases, the greybody factor radiation of bosons and fermions decreases significantly.

Finally, the interactions between BHs and fundamental fields help us understand BHs better.  We hope to investigate the Dirac and Klein-Gordon equations, as well as the Hawking radiation of the rotating modified Hayward BH [43] in the near future.

\bigskip
{\Large Acknowledgements}

I would like to thank the anonymous referees for their constructive comments and suggestions.
\newline

{\Large Data availability statement:}
My manuscript has no associated data.
\newline

{\Large Competing interests:} The authors declare there are no competing interests.
\bigskip

{\Large References}

[1] J. Bardeen, Proceedings of GR5, Tiflis, U.S.S.R. (1968).

[2] A. Borde, Phys. Rev. D 50, 3392 (1994).

[3] A. Borde, Phys. Rev. D 55, 7615 (1997).

[4] E. Ayon-Beato, A. Garc\i a, Phys. Rev. Lett. 80, 5056 (1998).

[5] K. Bronnikov, Phys. Rev. Lett. 85, 4641 (2000).

[6] K. Bronnikov, Phys. Rev. D 63, 044005 (2001).

[7] K.A. Bronnikov, V.N. Melnikov, G.N. Shikin, K.P. Staniukovich, Ann.Phys. (USA) 118, 84 (1979).

[8] S.A. Hayward, Phys. Rev. Lett. 96, 031103 (2006).

[9] Lin, K., Li, J., Yang, S.Z. Int. J. Theor. Phys. 52, 3771-3778 (2013).

[10] Abbas, G., Sabiullah, U. Astrophys. Space Sci. 352, 769-774 (2014).

[11] Amir,M., Ghosh, S.G. JHEP 1507, 015 (2015).

[12] Amir,M. Eur. Phys. J. C 76, 532 (2016).

[13] Angel Rincon, Victor Santos, Eur. Phys. J. C 80:910 (2020). 

[14] C. Bambi, L. Modesto, Phys. Lett. B 721, 329 (2013).

[15] T. De Lorenzo, C. Pacilioy et al, arXiv:1412.6015v1.

[16] P. Boonserm, T. Ngampitipan, P.Wongjun, Eur. Phys. J. C 79, 330 (2019).

[17] Ángel Rincón1 and  Victor Santos, Eur. Phys. J. C 80:910 (2020).

[18]  Ziou Yang and Yen-Kheng Lim, Phys. Rev. D 105, 124045 (2022).

[19] I. Sakalli, Phys. Rev. D 94, 084040 (2016). arXiv:1606.00896 [grqc].

[20] A. Al-Badawi, I. Sakalli, S. Kanzi, Ann. Phys. 412, 168026 (2020).

[21] A. Al-Badawi, S. Kanzi, I. Sakalli, Eur. Phys. J. Plus 135, 219 (2020).

[22] H. Gursel, I. Sakalli, Eur. Phys. J. C 80, 234 (2020).

[23] S. Kanzi, S.H.Mazharimousavi, I. Sakalli, Ann. Phys. 422, 168301 (2020).

[24] T. Harmark, J. Natario, and R. Schiappa, Adv. Theor. Math. Phys. 14, 727 (2010).

[25] I. Sakalli, Int. J. Mod. Phys. A 26, 2263 (2011); [erratum: Int. J. Mod. Phys. A 28, 1392002 (2013)].

[26] A. Al-Badawi, S. Kanzi, and ˙I. Sakallı, Eur. Phys. J. Plus 137, 94 (2022).

[27] S. Kanzi and I. Sakallı, Eur. Phys. J. C 81, 501 (2021).

[28] I. Sakallı and S. Kanzi, Annals Phys. 439, 168803 (2022).

[29] S. Kanzi and I, Sakallı, European Physical Journal Plus 137, 14 (2022).

[30] Sharif, M., Ama-Tul-Mughani, Q. Eur. Phys. J. Plus 134, 616 (2019).

[31] Petarpa Boonserm and Matt Visser, Phys.Rev.D78:101502, (2008).

[32] E. Newman, R. Penrose, J. Math. Phys. 3, 566 (1962).

[33] S. Chandrasekhar, The Mathematical Theory of Black Holes Clarendon, P. 543 London (1983).

[34] M. Abramowitz, I. A. Stegun (ed.), Handbook of Mathematical Functions
(Dover, New York, 1970).

[35] D.N. Page, Phys. Rev. D 13, 198 (1976).

[36] WG Unruh, Physical Review D 14, 3251 (1976).

[37] R. Jorge, E.S. de Oliveira, J.V. Rocha, Class. Quantum Gravity 32,
065008 (2015).

[38] J. Abedi, H. Arfaei, Class. Quantum Gravity 31, 195005 (2014).

[39] Y.G. Miao, Z.M. Xu, Phys. Lett. B 772, 542 (2017).

[40] I. Sakalli and S. Kanzi, Turk. J. Phys. 46, 51 (2022).

[41] M. Visser, Phys. Rev. A 59, 427438 (1999). arXiv: quant-ph/9901030.

[42] J. Lpez, J. Comput. Appl. Math. 102, 181 (1999). 

[43] C. Bambi and L. Modesto, Phys. Lett. B 721, 329 (2013).

\end{document}